\documentclass[utf8]{FrontiersinHarvard}
\usepackage{url,hyperref,microtype,subcaption}
\usepackage[onehalfspacing]{setspace}
\graphicspath{{fig-jpeg/}}

\usepackage{graphicx}
\usepackage{epstopdf, epsfig}
\usepackage{natbib}
\usepackage{amsmath}
\usepackage{amssymb}
\usepackage{amsbsy}
\usepackage{color}
\usepackage{calrsfs}
\usepackage{amsmath}
\usepackage{amsfonts, amsmath,mathrsfs}

\DeclareMathAlphabet{\mathbfsf}{\encodingdefault}{\sfdefault}{bx}{n}
\newcommand{\tens}[1]{\mathbfsf{#1}}
\definecolor{n_blue}{RGB}{52, 152, 219}
\definecolor{n_green}{RGB}{52,219,119}

\def \bfu {{\bf u}}

\def \Ga  {\mbox{Ga}}

\def \At  {\mbox{At}}
\def \Bo  {\mbox{Bo}}

\usepackage[normalem]{ulem}
\newcommand{\beq}{\begin{equation}}
\newcommand{\eeq}{\end{equation}}

\newcommand{\REM}[1]{{}}

\def\keyFont{\fontsize{8}{11}\helveticabold }

\def\firstAuthorLast{Ramadugu {et~al.}} 

\def\Authors{Rashmi Ramadugu\,$^{1,\dagger}$, Vikash Pandey\,$^{2,\dagger}$ and Prasad Perlekar\,$^{3,*}$}
\def\Address{$^{1}$ Sankhyasutra Labs, Bengaluru, India \\
$^{2}$ Nordita, KTH Royal Institute of Technology and
Stockholm University, Hannes Alfv{\'e}ns v\"ag 12, 10691 Stockholm, Sweden \\
$^{3}$ Tata Institute of Fundamental Research,  Gopanpally, Hyderabad 500046, India. \\
$\dagger$ These authors contributed equally to this work and share first authorship}
\def\corrAuthor{Corresponding Author: Prasad Perlekar}
\def\corrEmail{perlekar@tifrh.res.in}

\begin{document}

\onecolumn
\firstpage{1}

\title[Buoyancy-driven bubbly flows in Hele-Shaw cell]{Energy spectra of buoyancy-driven bubbly flow in a vertical Hele-Shaw cell} 

\author[\firstAuthorLast ]{\Authors} 
\address{} 
\correspondance{} 

\extraAuth{}

\maketitle

\begin{abstract}
 We present direct numerical simulations (DNS) study of confined buoyancy-driven
 bubbly flows in a Hele-Shaw setup. We investigate the spectral properties of the
 flow and make comparisons with experiments.  The energy spectrum obtained from
 the gap-averaged velocity field shows  $E(k)\sim k$ for $k<k_d$, $E(k)\sim
 k^{-5}$ for $k>k_d$, and an intermediate scaling range with $E(k)\sim k^{-3}$
 around $k\sim k_d$. We perform an energy budget analysis to understand the
 dominant balances and explain the observed scaling behavior.  We also show that
 the Navier-Stokes equation with a linear drag can be used to approximate large
 scale flow properties of bubbly Hele-Shaw flow.
 \keyFont{ \section{Keywords:} Buoyancy driven bubbly flows, Hele-Shaw flows}
\end{abstract}

\section{Introduction}

Flows generated by dilute bubble suspensions (bubbly flows) are relevant in many
natural and industrial processes \citep{clift1978}. As the bubbles rise due to
buoyancy and stir the fluid, they generate complex spatiotemporal flow structures
``pseudo-turbulence" \citep{lance_1991,riss18,mudde_rev_2005,mat20,pan20}. The underlying
physical mechanisms responsible for the flow are the interaction between wakes
caused by individual bubbles and the interaction of bubbles with the flow generated
by their neighbors \citep{riss18,mat20}.

Early experiments characterized pseudo-turbulence in bubbly flows at a low-volume
fraction by measuring the energy spectrum $E(k)\sim k^{-3}$ (where $k$is the wave
number). They argued that the power-law scaling appears due to a balance of energy
production with viscous dissipation \citep{lance_1991}. Subsequent experimental
studies have verified the power-law scaling in the energy spectrum
\citep{risso_legendre_2010,prakash2016energy,mendez,mat20}.

Only recent numerical studies have started investigating pseudo-turbulence at
experimentally relevant parameter ranges \citep{pan20,inno2021,pan22}. A
scale-by-scale energy budget analysis has unraveled the details of the energy
transfer mechanism. Buoyancy injects energy at scales comparable to the bubble
diameter; it is then transferred to smaller scales by nonlinear fluxes due to
surface tension and kinetic energy, where it gets dissipated by viscosity. Quite
remarkably, these studies also reveal that the statistics of the velocity
fluctuations do not depend either on the viscosity or density contrast
\citep{pan20,ramadugu2020,pan22}.

How does the physics of bubbly flows altered in the presence of confinement? Earlier
studies have investigated this question in a Hele-Shaw setup with bubbles whose
unconfined diameter is larger than the confinement width
\citep{risso_billet_2012,bouche2012homogeneous,bouche2014}. Numerical simulations and
experiments \citep{clift1978,kelley1997path,wang2014experimental,fil15}
on an isolated rising bubble show that, compared to an unconfined bubble, the wake
flow of the confined bubble is severely attenuated. Nevertheless, the experiments on
bubbly flows in the Hele-Shaw setup still observe the power-law scaling of
pseudo-turbulence between scales comparable to the bubble diameter and twenty times
the bubble diameter. 

In this paper, we perform a numerical investigation of buoyancy-driven bubbly flow
in a Hele-Shaw setup. To make a  comparison with experiments, we choose moderate volume
fractions $\phi=5-10\%$. We investigate the energy spectrum of the gap-averaged
velocity field and, consistent with experiments, observe an interediate power-law 
scaling in the energy spectrum $E(k)\sim k^{-3}$. Using a scale-by-scale energy budget analysis, we
show that confinement dramatically alters the energy budget compared to the
unbounded bubbly flows. The viscous drag due to the confining walls balances energy
injected by buoyancy at large scales. Nonlinear transfer mechanisms due to surface
tension and kinetic energy are negligible. Finally, we show that two-dimensional
Navier-Stokes equations with an added drag term can be used as a model to study
large scale flow properties.

The rest of the paper is organised as follows. In section \ref{sec:model}, we
discuss the governing equations and the details of the numerical method used. In
section \ref{Sec:multi_hs}, we present results for bubbly flows in the Hele-Shaw
setup  and study the energy budget. We then show that the two-dimensional Navier-Stokes equations
with a linear drag is a good model to study large scale properties of bubbly flows
under confinement. Finally, in section~\ref{Sec:conclusions}, we present our
conclusions.

\section{Equations and numerical methods}
 \label{sec:model}

\begin{figure}[!h]
    \centering
    \includegraphics[width=\linewidth]{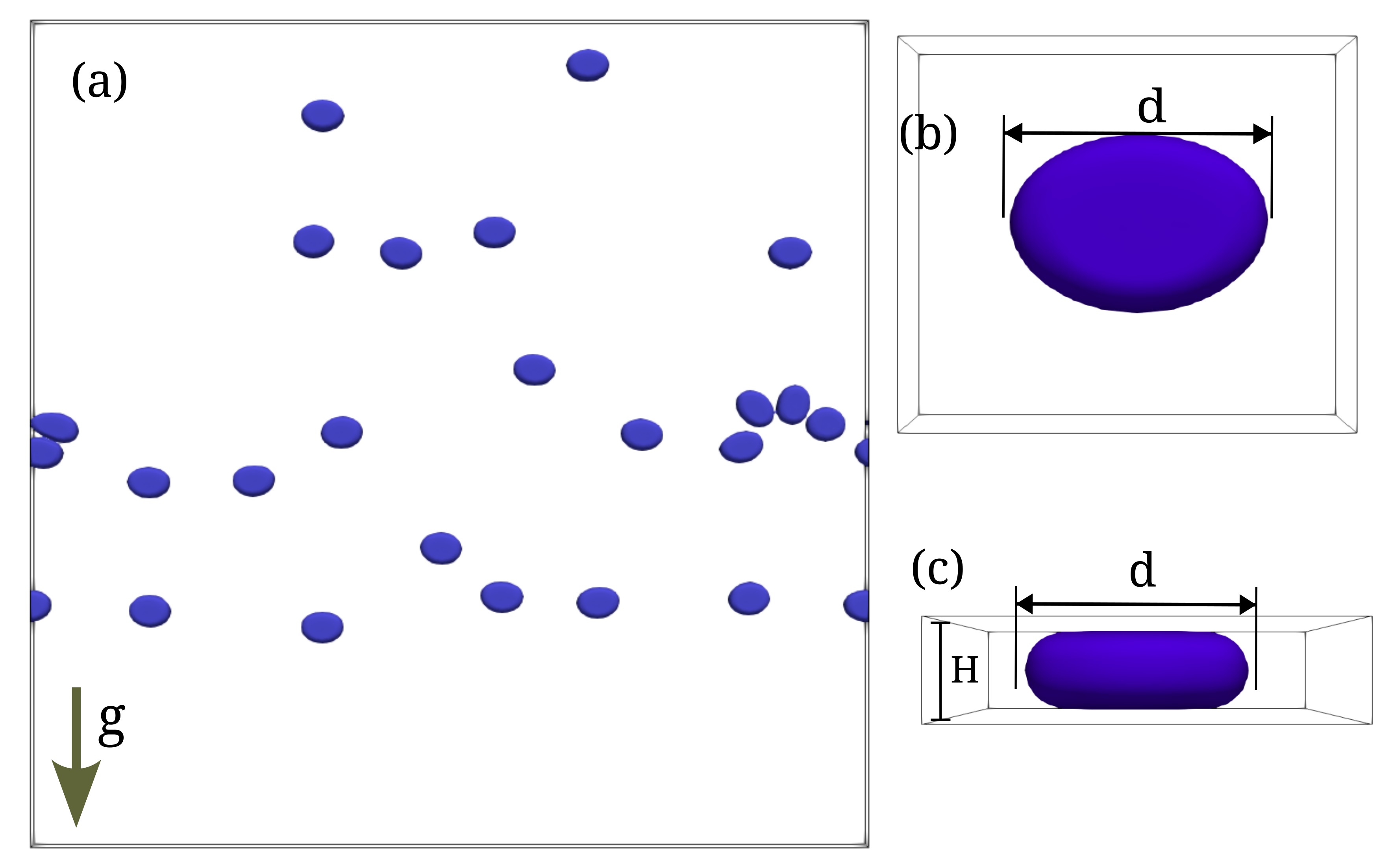}
    \caption{\label{fig:hs_sketch} (a) Representative plot showing showing bubbles of diameter $d$ in a Hele-Shaw setup. The length along $x-$ and $z-$ directions is $L$, and the  gap width in $y-$ direction is $H$; (b) Top view of a bubble (zoomed view); (c) Front  view of the bubble (zoomed view).}
\end{figure}

We study the dynamics of bubbly flows in a vertical Hele-Shaw cell (see
Fig.~\ref{fig:hs_sketch}) by solving the Navier-Stokes equations with surface tension
force acting at the interface 
\begin{eqnarray}
    \partial_t c  + {\bf v} \cdot \nabla^* c&=& 0, {\nabla}^* \cdot {\mathbf v} = 0,~\rm{and} \\
    \rho(c) (\partial_t +  \mathbf{v} \cdot {\nabla}^*) {\bf v} &=& {\nabla}^* \cdot \left[ \mu(c) (\nabla^* {\bf v} + \nabla^* {\bf v}^{T}) \right] - \nabla^* P + \mathbf{F}^g + \mathbf{F}^{\sigma}\label{3d_NS}.
\end{eqnarray}
Here, $\nabla^*\equiv(\partial_x,\partial_y,\partial_z)$, $c$ is an indicator
function whose value is $0$ inside the bubble phase and $1$ in the fluid phase,
$\mathbf{ F}^{g} \equiv  [\rho_a-\rho(c)] g\hat{\mathbf{e}}_z$ is the buoyancy
force, $\mathbf{v} = (v_x,v_y,v_z)$ is the hydrodynamic velocity, $P$ is the
pressure, the local density $\rho(c)\equiv \rho_1 c + \rho_2 (1-c)$, the local
viscosity $\mu(c)\equiv \mu_1 c + \mu_2 (1-c)$, $\rho_2$ ($\rho_1$)    is  the
bubble (fluid) density, $\mu_2$ ($\mu_1$) is the bubble (fluid) viscosity, and
$\mathbf{ F}^{\sigma} \equiv \sigma \kappa \nabla c$ is the surface tension force at
the interface
\citep{brackbill_1992} with $\sigma$ as the coefficient of surface tension and 
$\kappa$ the interface curvature. The bubble volume fraction $\phi \equiv[\int (1-c)
d\mathbf{ x}]/(L^2H)$, where $L$ is the length along the $x-$ and $z-$ directions,
and $H$ is the gap width between the two parallel plates of the Hele-Shaw cell. In
what follows, $\rho_1$ ($\mu_1$) denotes the density (viscosity) of the liquid
phase, and  $\rho_2$ ($\mu_2$) denotes the density (viscosity) of the bubble phase.

The non-dimensional numbers that characterize the flow are the Galilei number $\Ga
\equiv {\rho_1 \sqrt{\delta \rho g d}d}/{\mu_1}$, the Bond number $\Bo \equiv {\delta \rho g
d^2}/{\sigma}$, and the Atwood number $\At \equiv {\delta \rho}/{(\rho_1 +
\rho_2)}$ with $\delta \rho = (\rho_1 - \rho_2)$.  For brevity, in the following
sections, we will refer to Eq.~\eqref{3d_NS} as NSHS.

\subsection{Gap width averaged equations:}
Experiments often use gap width averaged velocities to study statistical properties
of the flow. Following the procedure outline in
\citep{gondret1997,alexakis2018cascades}, and by assuming density $\rho$ to be
constant along wall-normal direction, we get the following equations for the
averaged in-plane horizontal components of the velocity:

\begin{eqnarray}
    (\partial_t + \bfu \cdot \nabla) \overline{c} &=& 0,~\rm{and}~  \nabla \cdot \bfu = 0, \\
    \overline{\rho} (\partial_t + \bfu \cdot \nabla) \bfu &=& -  \overline{\rho} \nabla \cdot  \overline{{\bf v}^\prime {\bf v}^\prime} + \nabla \cdot \left( 2 {{\mu} \tens{S}} \right) +  \mathbf{\overline{F}}^{d} - \nabla \overline{P} + \mathbf{\overline{F}}^g + \mathbf{\overline{F}}^{\sigma}.  \label{eq:NS_2d_HS}
\end{eqnarray}
Here, $\overline{(.)}\equiv (1/H) \int_0^H (.) dy$ denotes gap averaging,
$\nabla\equiv(\partial_x,\partial_z)$,   $\bfu\equiv (u_x(x,z),u_z(x,z))$  is the
gap averaged velocity field with $u_x=\overline{v}_x$, $u_z =\overline{v}_z$,  ${\bf
v}^\prime({\bf x})=(v_x-u_x,v_z-u_z)$ are the three-dimensional residual velocity
fluctuations, $\overline{P}(x,z)$ is the gap-averaged pressure field,
$\tens{S}=\nabla {\bfu} + \nabla \bfu^{T}$ is the gap-averaged strain-rate tensor,
$\overline{\rho}(x,z)$ is the density field, ${\overline{\bf{F}}}^g=[\rho_a -
\overline{\rho}]g \hat{z}$ is the buoyancy force, $\overline{\bf{F}}^\sigma$ is the
surface tension force. The viscous dissipation contributes in two parts: (a)
small-scale dissipation $\nabla \cdot \left( 2 \overline{{\mu}} \tens{S} \right)$,
and (b) viscous drag due to walls  ${\overline{\bf F}}^d  = [\mu (\nabla {\bf v} +
\nabla {\bf v}^T) \cdot \hat{y}]^{H}_0$.

\subsection{Numerical Method}    
We use a second-order finite-volume solver PARIS \citep{aniszewski2021parallel} to
simulate NSHS (Eq. \eqref{3d_NS}). For bubble tracking PARIS employs a front
tracking method, and the time marching is performed either using the first order
Euler method or the second order Crank-Nicolson method. 

\subsection{Initial conditions and parameters}
We consider a cuboid of breadth $L_y=H$, and with equal length and height
($L_x=L_z=L$) [see Fig.~\ref{fig:hs_sketch}].  We use periodic boundary conditions
in the $x$ and $z$ directions, and impose no-slip velocity boundary ${\mathbf u}=0$
at the walls  ($y=0$ and $y=H$). We place $N_b$ bubbles in random positions and
initialize each one as an ellipsoid of volume $V=4.73\times10^3$ (mono-disperse
suspension). The bubbles  are allowed to relax in the absence of gravity until they
achieve the equilibrium pan-cake-like configuration \citep{gan20} with diameter
$d/H=2$. In table~\ref{tab:runs_multi_HS3d} we summarize the parameters used in
our simulations.

\begin{table*}[!h]
\caption{\label{tab:runs_multi_HS3d}
Parameters used in our simulations.  We fix $L=512$, $H=12$, $N_x=N_z$, and $d=24$
for all the runs.}
\begin{center}
\begin{tabular}{cccccccccccc}
 \hline\noalign{\smallskip}
         \# &$N_b$ & $N_x$ & $N_y$ &$ \rho_1$  & $\mu_1$ & $\mu_1/\mu_2$ & $\Ga$    &  $\Bo$      & $\At$ & $\phi$\\
       \noalign{\smallskip}\hline\noalign{\smallskip}
      $\tt{H1}$     & $24$ & $1024$ & $24$ & $1.0$ & $0.16$ & $1$  & $294$ & $1.8$ & $0.08$ & $0.0552$ \\
      $\tt{H2}$     & $46$ & $512$ & $32$ & $1.0$ & $0.16$ & $1$  & $294$ & $1.8$ & $0.08$ & $0.1058$ \\
      $\tt{H3}$     & $24$ & $512$ & $32$ & $1.0$ & $0.42$ &  $20$ & $274$ &  $3.4$  & $0.9$ & $0.0552$\\
\noalign{\smallskip}\hline
     \end{tabular}
     \end{center}
\end{table*}

\section{Results}
\label{Sec:multi_hs}
\begin{figure}[!h]
    \centering
    \includegraphics[width=0.6\linewidth]{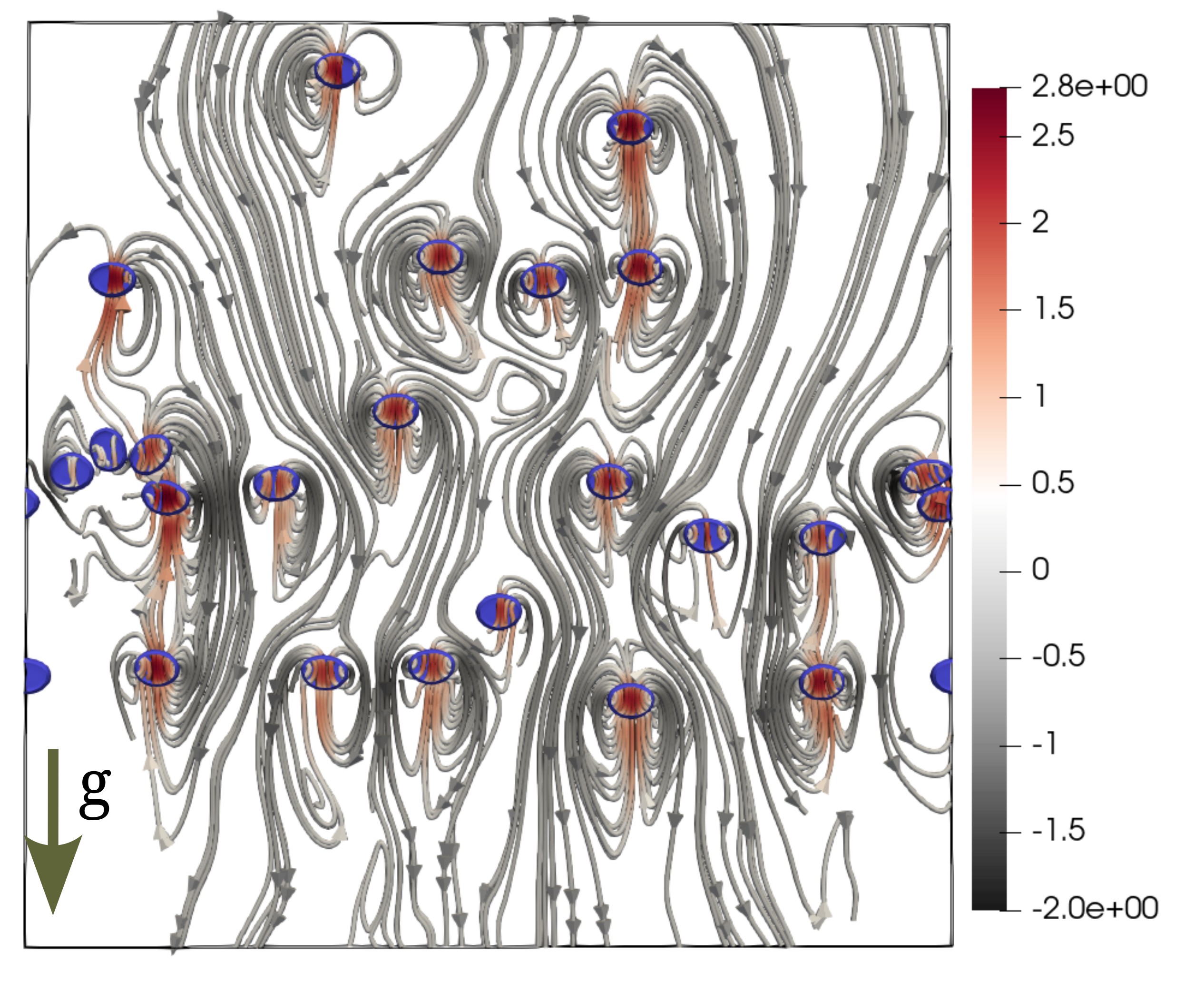}
    \caption{\label{fig:bubb_stream} Instantaneous bubble configuration superimposed
    with flow streamlines in the steady-state (run ${\tt H1}$). The streamlines are
colored according to the z-component of the velocity.}
\end{figure}

In this section, we present the results of our numerical investigations.  We monitor
the time evolution of the gap-averaged energy and investigate the corresponding flow
properties in a statistically steady state. The plot in Fig.~\ref{fig:bubb_stream}
shows a typical snapshot of the bubble configuration along with the flow streamlines
in the steady-state. Similar to the experiments \citep{bouche2014}, we observe that
the flow disturbances are mostly localized in the bubble vicinity. Furthermore, the
horizontal alignment of bubbles is also observed in experiments \citep{bouche2014}
as well as numerical simulation of stratified bubbly flows in a Hele-Shaw setup
\citep{gan20}. As is conventional in the experiments
\citep{bouche2012homogeneous,bouche2014}, we also investigate the spectral
properties of the gap-averaged velocity field  \eqref{eq:NS_2d_HS}. 

\subsection{Time evolution}
From \eqref{eq:NS_2d_HS}, we obtain the following balance equation for the gap-averaged kinetic energy $E$
\begin{equation}
\partial_t \underbrace{\langle \frac{\overline{\rho}~\bfu^2}{2} \rangle}_{E}
 = -\underbrace{ 2 \langle \overline{\mu} \mathcal{\tens{S}}:\mathcal{\tens{S}} \rangle}_{\epsilon_\mu} 
  + \underbrace{\langle [\rho_a -\overline{\rho}] u_z g \rangle}_{\epsilon_{inj}} + 
  \underbrace{\langle \mathbf{\overline{F}}^\sigma\cdot \bfu \rangle}_{\epsilon_\sigma} + 
  \underbrace{\langle \mathbf{\overline{F}}^d\cdot \bfu \rangle}_{\epsilon_d}, 
\end{equation}
where $\epsilon_\mu$ is the gap-averaged viscous energy dissipation, $\epsilon_d$ is the dissipation due to drag, $\epsilon_{inj}$ is the gap-averaged energy injected due to buoyancy, $\epsilon_{\sigma}$ is the contribution due to the surface tension, and the angular brackets denote spatial averaging.

\begin{figure}
    \centering
    \includegraphics[scale=0.8]{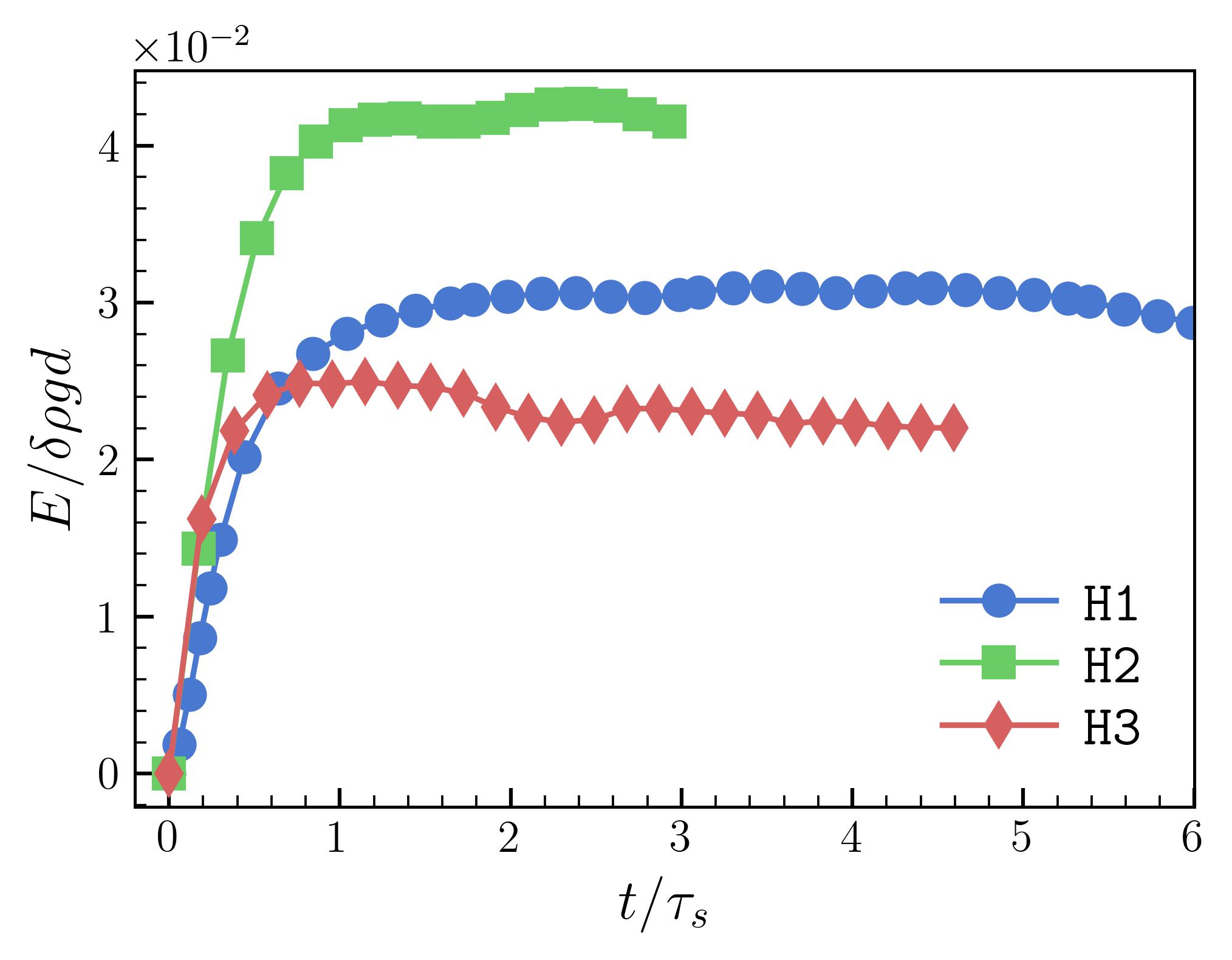}
     \caption{\label{fig:ke_check} Time evolution of the kinetic energy $E$. A
         steady-state is attained for $t\geq 0.8\tau_s$, where $\tau_s =
     L/\sqrt{gd}$.}
\end{figure}

In Fig.~\ref{fig:ke_check}, we plot the time-evolution of the  kinetic energy $E$
and observe that a statistically steady state is achieved  for $t>0.8\tau_s$.
Furthermore, in table~\ref{tab:multi_HSenbal} we show that the energy injected by
buoyancy is primarily balanced by the dissipation due to drag in the steady state
($\partial_t E\approx 0$).

\begin{table*}
\caption{\label{tab:multi_HSenbal} Time-averaged values of the energy injection $\epsilon_{inj}$, viscous dissipation $\epsilon_\mu$, and dissipation due to drag $\epsilon_d$ in the statistically steady state.}
 \begin{center}
\begin{tabular}{cccc}
 \hline\noalign{\smallskip}
 \#   &  $\epsilon_{\mu}\times10^{-3}$    &  $-\epsilon_{d}\times10^{-3}$      & $\epsilon_{inj}\times10^{-3}$   \\
       \noalign{\smallskip}\hline\noalign{\smallskip}
      $\tt{H1}$   &   $0.8$ & $5.6$ & $6.5$  \\
      $\tt{H2}$    &  $1.1$ &  $11.6$ & $11.6$\\
      $\tt{H3}$     & $0.8$ &  $8.8$ & $ 9.1 $\\
 \noalign{\smallskip}\hline
     \end{tabular}
     \end{center}
\end{table*}

\subsection{Energy spectra and scale-by-scale energy budget}
 The energy spectrum and co-spectra for the gap-averaged velocity field are defined as:
\begin{eqnarray}
    E(k) &\equiv & \sum_{k-1/2<m<k+1/2} |\hat{\mathbf{u}}({\bf m})|^2, \nonumber \\
 E^{\rho u} (k) &\equiv & \sum_{k-1/2<m<k+1/2} \Re[\hat{(\rho \mathbf{  u})}({-\bf
 m}) \hat{\mathbf{u}}({\bf m})]. \nonumber
\end{eqnarray}
Here, $\hat{(\cdot)}$ denotes the Fourier transformed fields.

\begin{figure}[!h]
    \centering
    \includegraphics[width=0.45\textwidth]{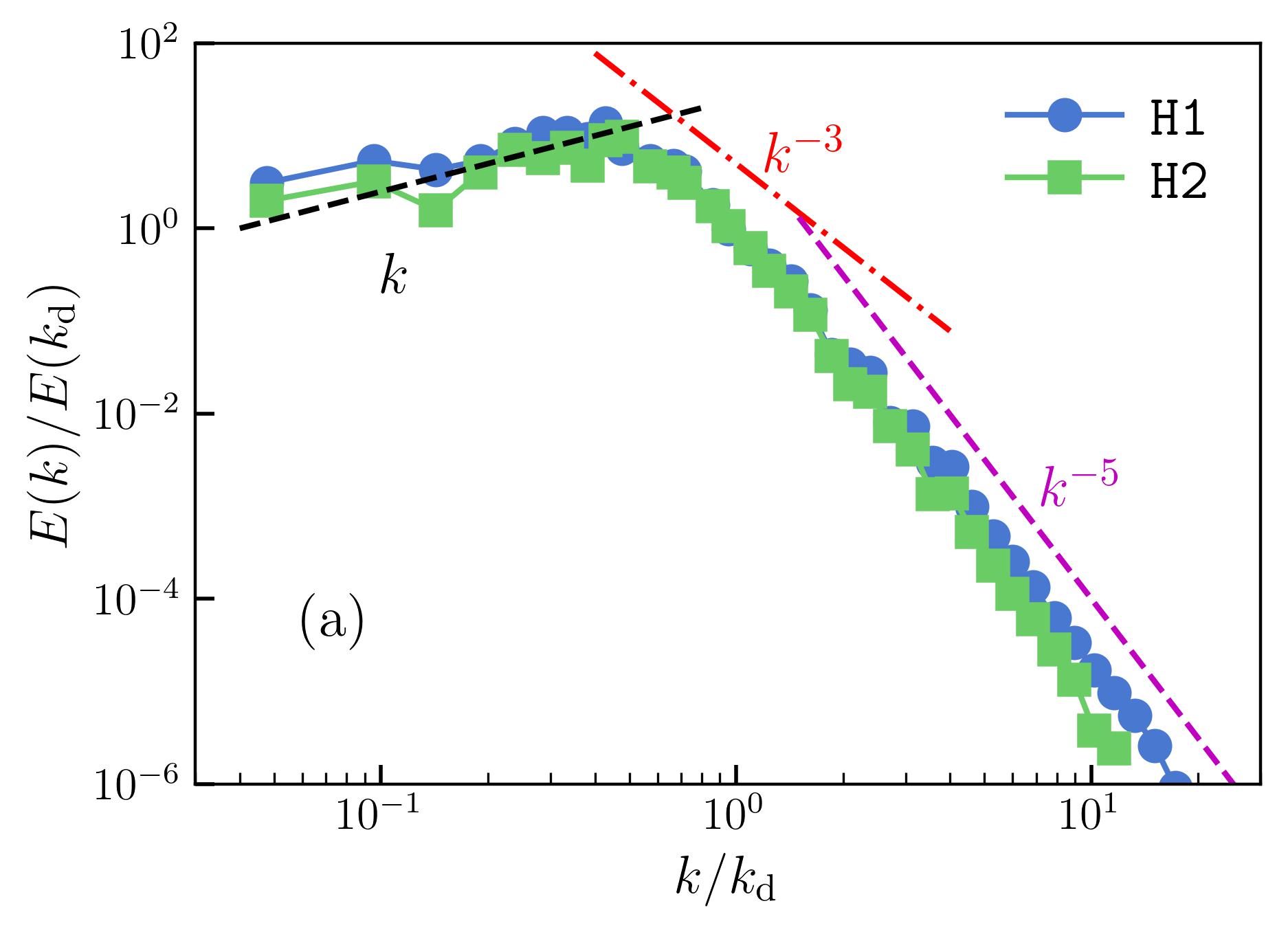} 
    \includegraphics[width=0.45\textwidth]{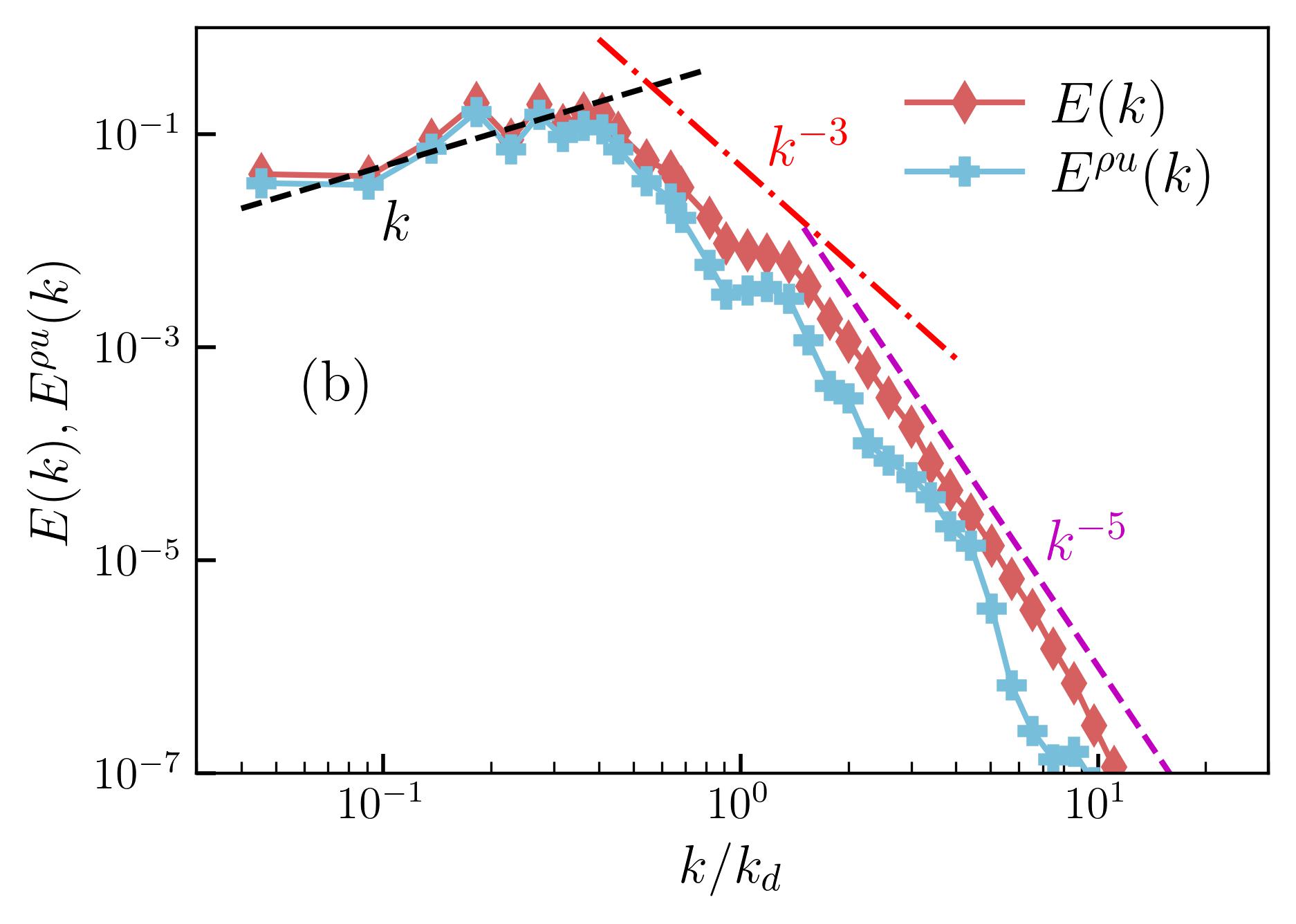} 
    \caption{(a) Log-log of the energy spectra ($E(k)$ versus $k$)  for low $\At$ runs
        ${\tt H1}$ and ${\tt H2}$.  (b) Log-log plot of the energy spectra ($E(k)$)
        and cospectra ($E^{\rho u}(k)$) for high $\At=0.9$ run $H3$ ($\Ga=274$, $\phi=0.05$). \label{fig:spectra_hs_2d}}
\end{figure}

In Fig.~\ref{fig:spectra_hs_2d},  we plot the energy spectra $E(k)$ and cospectra
$E^{\rho u}(k)$ for our simulations ${\tt H1-H3}$ \footnote{As the density contrast is negligible for the low $\At$, we do not plot the co-spectra for ${\tt H1,H2}$.}.  From the plots, we can identify different scaling regimes: (a) For $k<<k_d$ we observe $E(k)\sim k$, where $k_d$ is the wavenumber corresponding to the bubble diameter;  (b) Around $k \sim k_d$, we find a short $-3$ scaling regime followed by a steeper decay of the spectrum. Our simulations are also consistent with earlier  experiments that also observe an intermediate  $k^{-3}$ scaling subrange for $0.2 \lessapprox k/k_d \lessapprox 1$ \citep{bouche2014}.

\cite{risso11} argues that the $k^{-3}$ scaling could be modelled as a signal consisting of a sum of localized random bursts. Although this explanation is consistent with Fig.~\ref{fig:bubb_stream},  it does not highlight the underlying mechanisms that generate the observed scaling. \cite{lance_1991} take an alternate viewpoint and argue that the balance of energy production and viscous dissipation leads to the $k^{-3}$ scaling.

The scaling of the energy spectrum we observe differs earlier studies on two-dimensional unbounded flows \citep{ramadugu2020,inno2021} at comparable $\Ga$. They find an inverse energy cascade with $E(k)\sim k^{-5/3}$ for $k<k_d$ and a $E(k)\sim k^{-3}$ scaling for $k>k_d$ due to the balance of energy injected by surface tension with viscous dissipation.

In what follows, we present an energy budget analysis to explain the observed scaling of the energy spectrum.

\subsubsection{Energy budget}
Since the scaling behaviour observed in our simulations ${\tt H1-H3}$ is identical, we perform the energy budget analysis using our highest resolution simulation ${\tt H1}$.  Ignoring inertia and assuming a statistically steady state, from \eqref{eq:NS_2d_HS} we get the following energy budget equation \citep{verma_2019,pope}:
\begin{equation}
 {\mathcal F}(k) + T^\sigma(k)  = \underbrace{\nu k^2 E(k)}_{D(k)} +  \mathcal{D}(k) ,
 \label{eq:budget}
\end{equation}

where $T(k)$ is the nonlinear kinetic energy transfer, $D(k)$ is the viscous
dissipation, $T^\sigma(k)=\sum^\prime \Re[\hat{\bf F}^\sigma({\bf m}) \cdot \hat{\bfu}(-{\bf
m})]$ is the nonlinear transfer due to surface tension, 
$\mathcal{D}(k)=-\sum^\prime \Re[\hat{\bf F}^d({\bf m}) \cdot \hat{\bfu}(-{\bf m})]$ is the
viscous dissipation due to drag, $\mathcal{F}(k)=\sum^\prime \Re[\hat{F}^g({\bf m})
\hat{u}_z(-{\bf m})]$ is the energy injection due to buoyancy. Here, $\sum^\prime \equiv
\sum_{|{\bf m}| < {k-1/2}}^{k+1/2}$ indicates summation over all wave-numbers in a
circular shell around wavenumber $k$.

\begin{figure}[!h]
\centering
\includegraphics[width=0.45\linewidth]{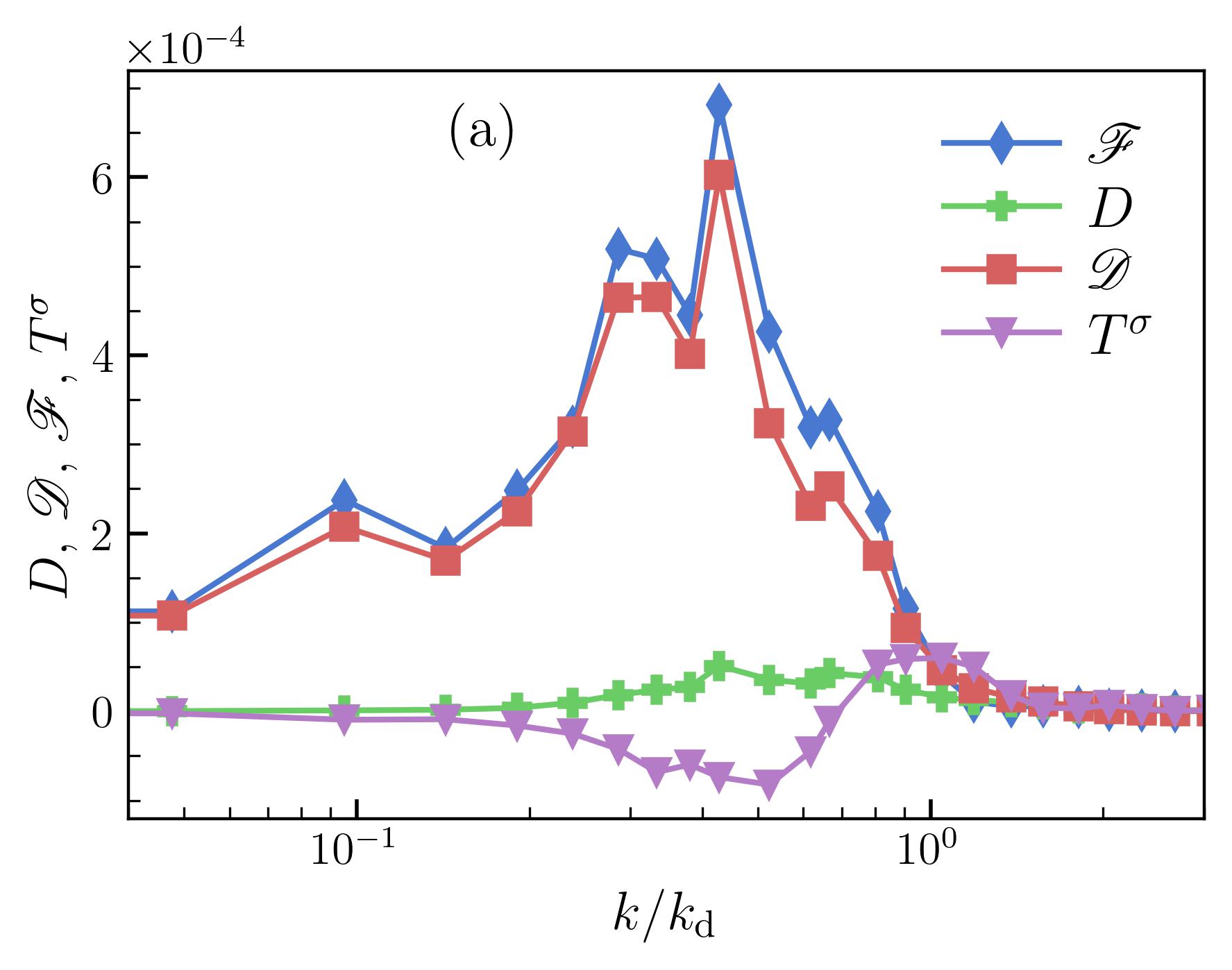}
\includegraphics[width=0.45\linewidth]{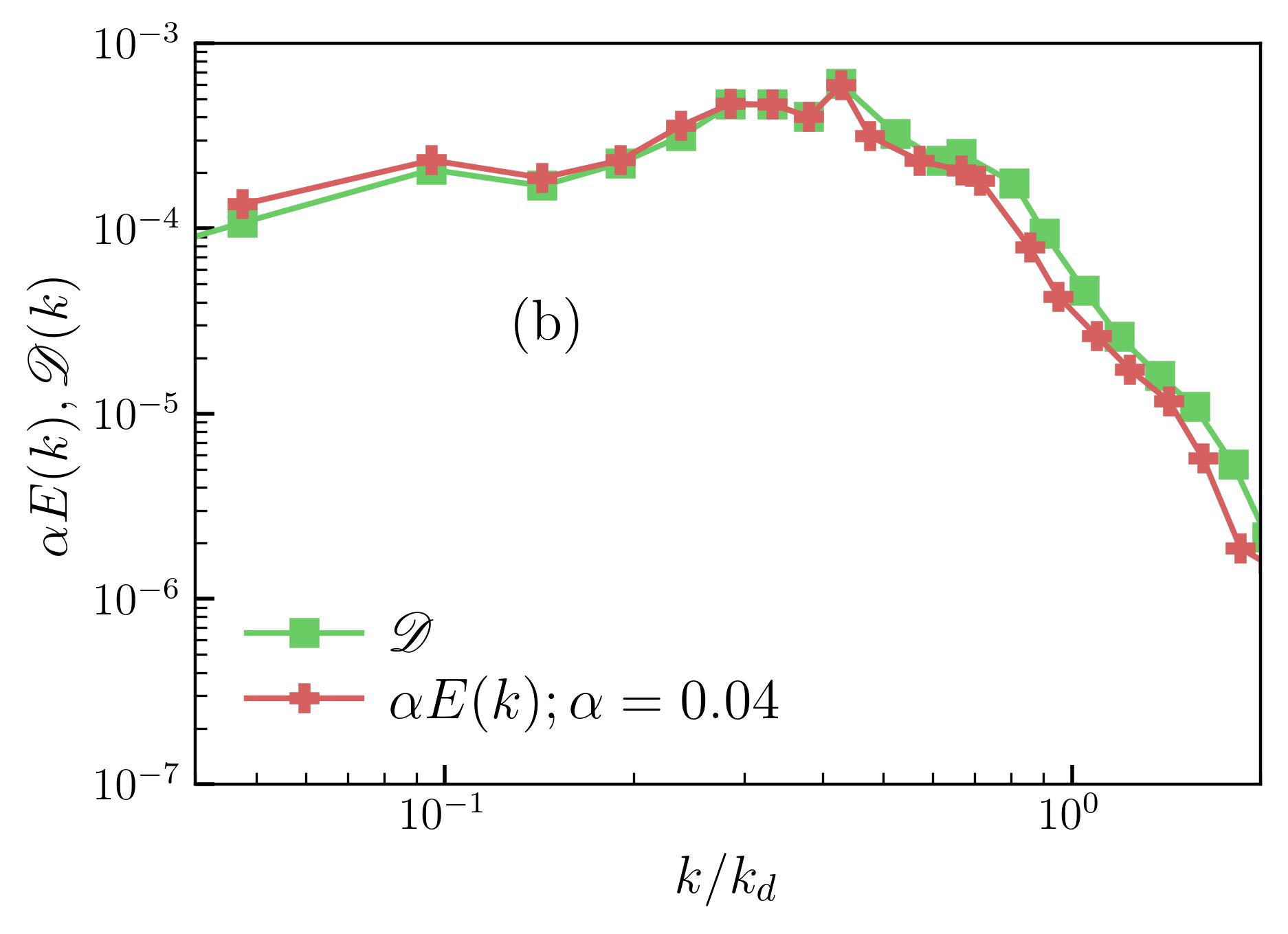}
\caption{\label{fig:balance} (a) Lin-log plot of the different contributions to the spectral energy budget \eqref{eq:budget} obtained from run ${\tt H1}$.  (b) Log-log plot showing comparison of the scaled energy spectrum $\alpha E(k)$ and  the dissipation due to drag ${\mathcal D}$(k) for $k/k_d < 2$.}
\end{figure}

The plot in Fig.~\ref{fig:balance}(a) shows the different contributions to the
budget.  Clearly for $k<k_d$, the energy injected by buoyancy is balanced by the
drag ($F^g(k)\sim  \mathcal{D}(k)$) and other contributions are subdominant. This
justifies our assumption of ignoring the inertial terms. In
Fig.~\ref{fig:balance}(b), we show that a linear drag approximation $F^d(k) \sim \alpha
E(k)$ (with $\alpha=0.04$) is in excellent agreement with ${\mathcal{D}}(k)$. Next
we approximate the energy injected by buoyancy as $F^g(k) \sim  \sqrt{E(k)
E_\rho(k)}$, where $E_\rho(k)=\sum^{\prime} |\hat{\rho}({\bf m})|^2$. Noting that for
$k<<k_d$, i.e. for scales much larger than bubble size, the density field can be
approximated by white noise $E_\rho(k)\sim k$ and by balancing the energy injected
by buoyancy with drag, we obtain $E(k)\sim k$. This explains the scaling observed in
our simulations for $k<k_d$.

 \begin{figure}[!h]
\centering
\includegraphics[width=0.32\linewidth]{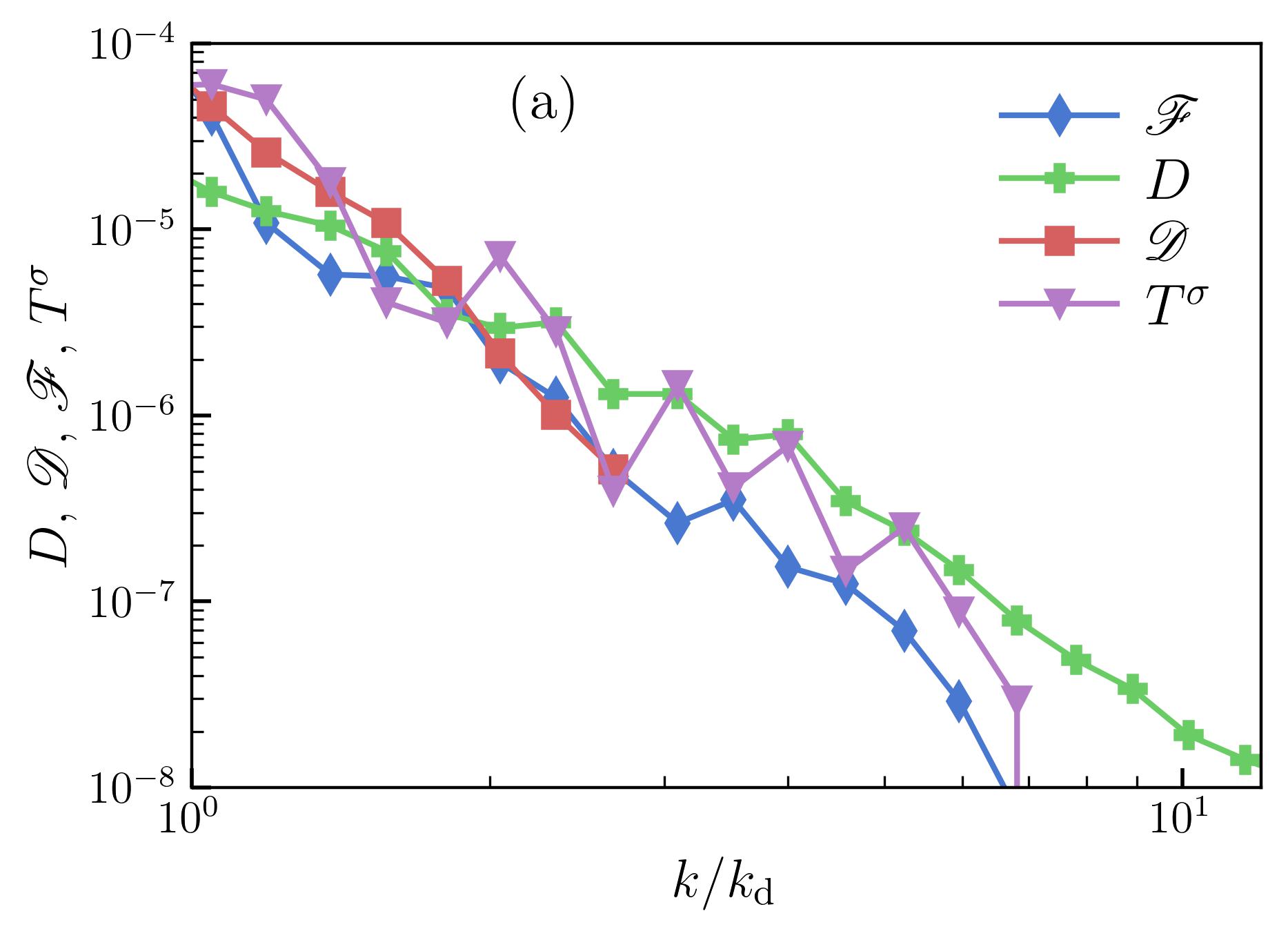}
\includegraphics[width=0.32\linewidth]{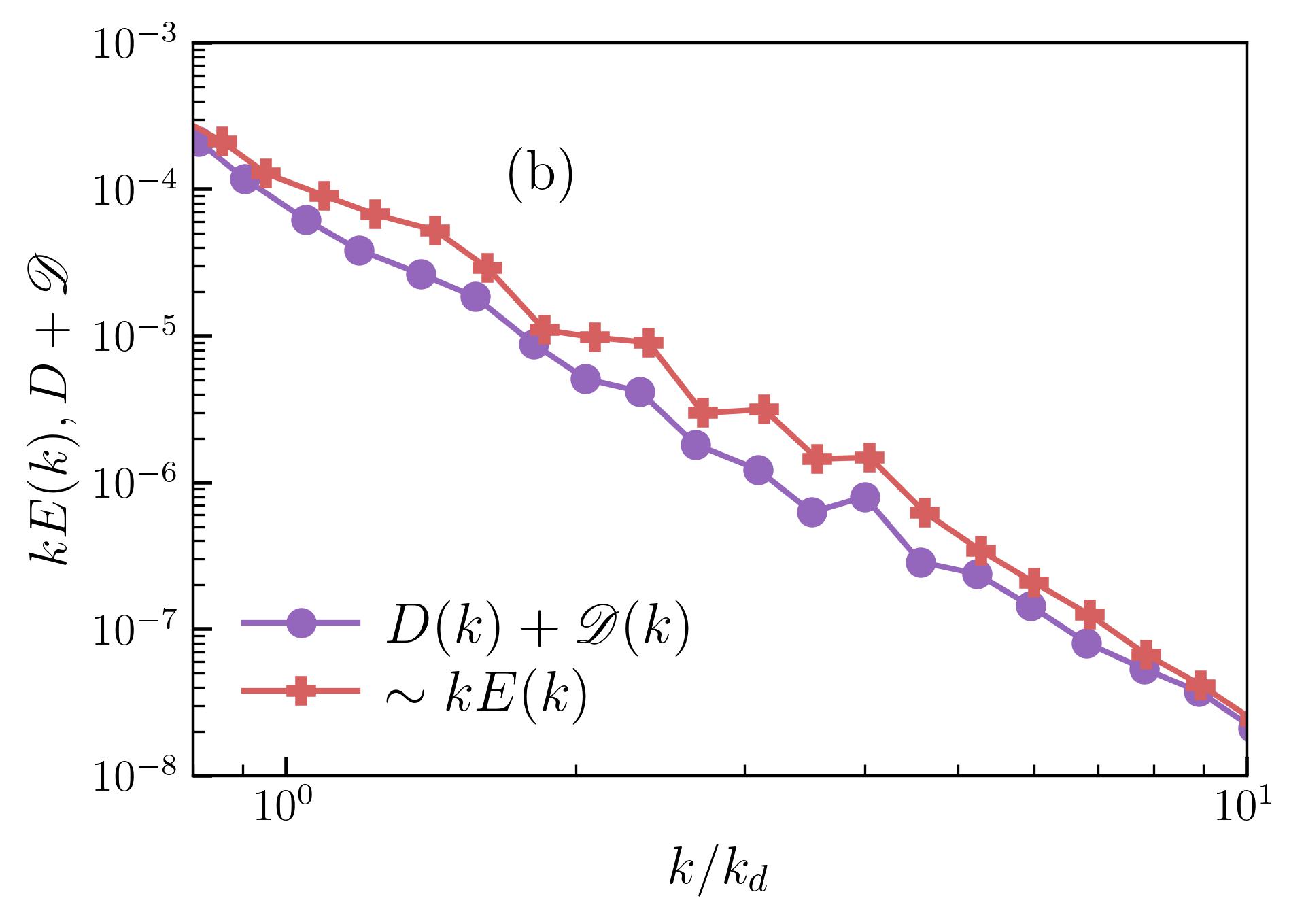}
\includegraphics[width=0.32\linewidth]{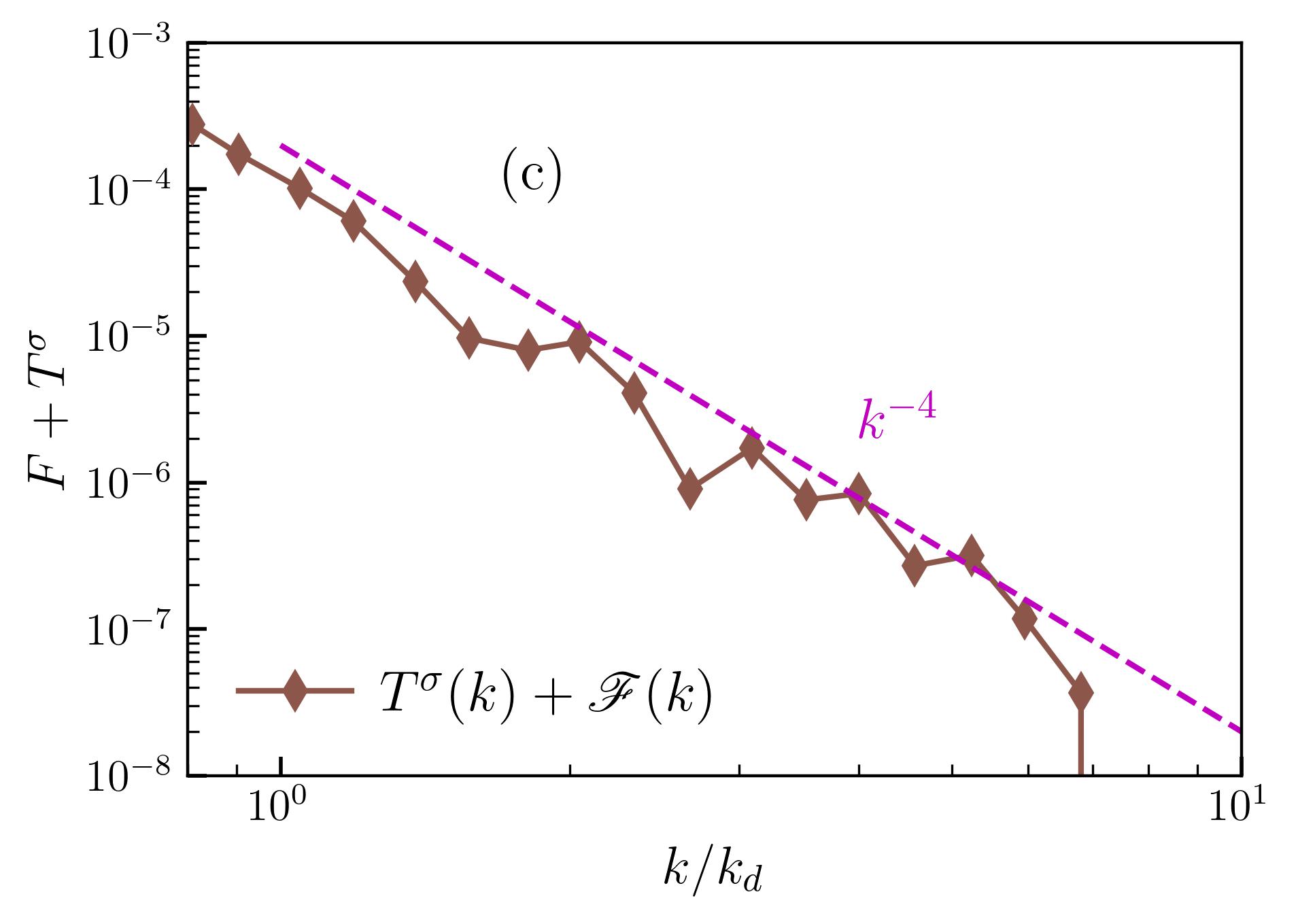}
\caption{\label{fig:balz}  (a) Zoomed in plot showing contributions to the spectral energy budget for $k>k_d$. (b) Log-log plot showing comparison of the scaled dissipation spectrum $\sim k E(k)$ and 
the net dissipation  $D(k)+{\mathcal D}(k)$ for $k> k_d$. (c) Log-log plot showing different scaling regimes in the net energy injection ${\mathcal F}(k)+ T^\sigma (k)$ for $k>k_d$.}
\end{figure}

The situation is more complicated for $k>k_d$. The zoomed-in plot of the energy
balance  (see Fig.~\ref{fig:balz}(a)) reveals that  both buoyancy and the surface
tension inject energy that gets dissipated by the viscous forces ($D+\mathcal{D}$),
and there is no dominant balance ($\mathcal{F} + T^\sigma \sim D + \mathcal{D}$). In
Fig.~\ref{fig:balz}(b), we show that the net dissipation $D+{\mathcal{D}}\sim  k
E(k)$. Similarly, the net production  $\mathcal{F} + T^{\sigma} \sim k^{-4}$ for $k
> k_d$. Therefore, by balancing the net injection with dissipation  we get $E(k)\sim
k^{-5}$ scaling for $k > k_d$. Given the limited cross-over scaling range $E(k)\sim
k^{-3}$  in Fig.~\ref{fig:spectra_hs_2d}, we are unable to argue about the
underlying mechanisms. Thus the plausible explanation for the $-3$ scaling is the
argument by \cite{risso11} that we have discussed in the previous section.

\subsection{Two-dimensional Navier-Stokes equations with a linear drag (NSD)}
In this section we investigate whether two-dimensional Navier-Stokes with a
linear drag coefficient \eqref{eq:drag} is able to model the confined bubbly flows.
In the following, we assume all the fields are two-dimensional and for comparison with the
gap-averaged quantities, we choose the same symbols.  
\begin{eqnarray}
    D_t c &=& 0,~\rm{and}~  \nabla\cdot {\bfu} = 0, \\
    \rho(c)D_t \mathbf{u} &=&  \nabla \cdot \left[ 2 \mu(c) \tens{S} \right] - \nabla P + {\bf F}^g + {\bf F}^{\sigma} - \alpha {\bf u}.  \label{eq:drag}
\end{eqnarray}
We perform the NSD simulations with a square domain of area $L^2$ and discretize it with $2048^2$ equi-spaced  points. The bubbles are initialized as circles of diameter $d=24$ and all the parameters of the simulation are identical to our run ${\tt H1}$ and we fix the drag coefficient $\alpha=0.04$. Our choice for the value of $\alpha$ is motivated by Fig.~\ref{fig:balance}. We use a front-tracking-pseudo-spectral method to evolve \eqref{eq:drag}.  For details of the numerical scheme, we refer the reader to \cite{ramadugu2020}. Below we discuss the statistical properties of the  flow in the steady state.

\begin{figure}[!h]
    \centering
    \includegraphics[width=0.48\textwidth]{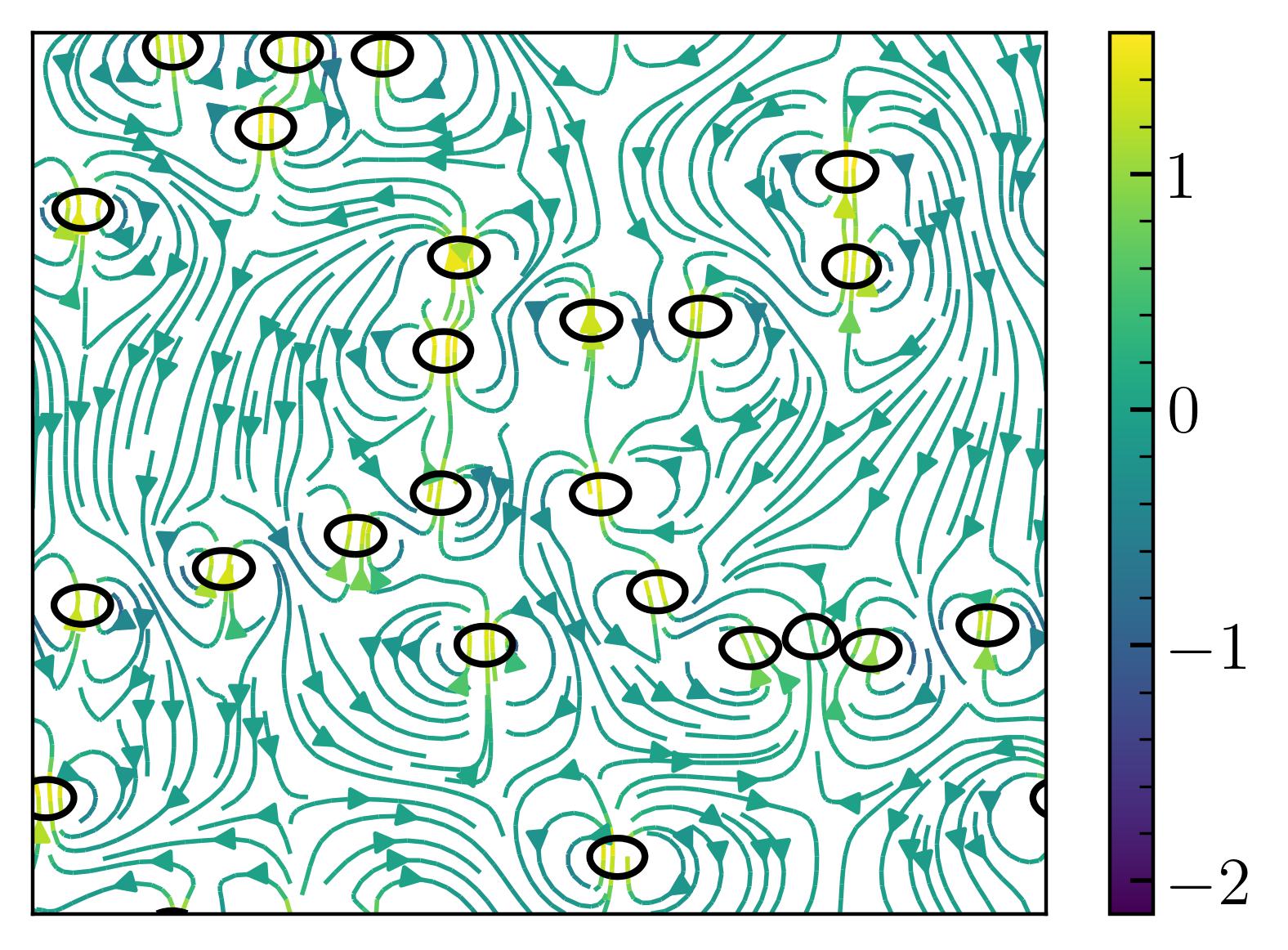}
    \put(-255,170){{\color{black}\bf ($a$)}}
    \includegraphics[width=0.48\textwidth]{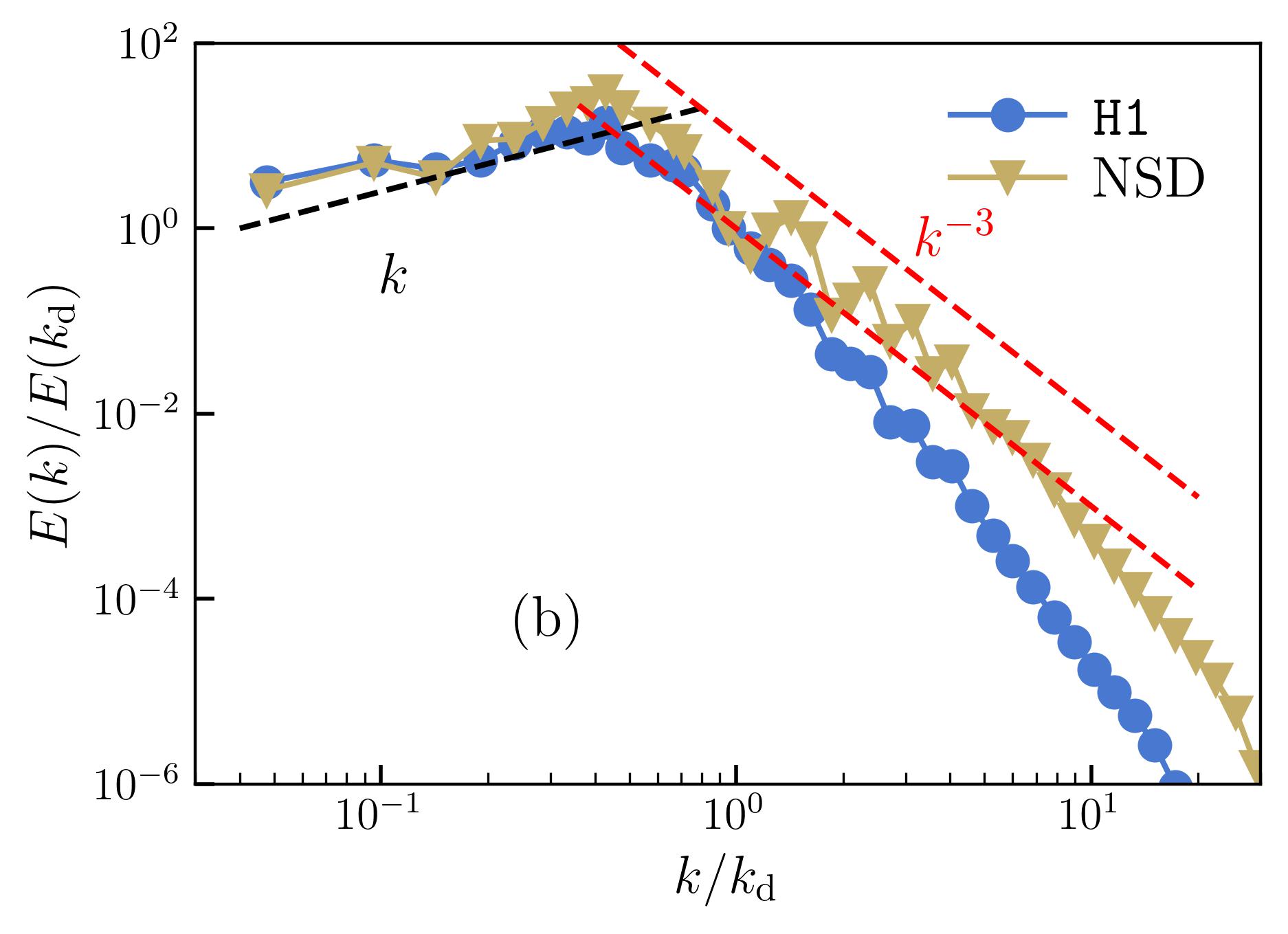}
    \caption{(a) Snapshot of the bubble positions overlaid with flow streamlines. (b) Comparison of the gap-averaged energy spectra for our NSHS run ${\tt H1}$  with the energy spectra obtained using simulation of the NSD equation \eqref{eq:drag}.    \label{fig:compsec}}
\end{figure}

In Fig.~\ref{fig:compsec}(a), we plot the bubble configuration and the flow streamlines. Clearly the large scale flow properties resemble those observed for the NSHS simulation. The flow disturbances are localized in the vicinity of the bubbles and we also observe horizontal alignment of bubbles.

The plot in Fig.~\ref{fig:compsec}(b) shows a comparison of the gap-averaged energy spectrum $E(k)$ 
obtained from the NSHS equation with that obtained from NSD equation \eqref{eq:drag}.
We find that the energy spectrum are nearly identical for $k<k_d$, $E(k)\sim k$. However, discrepancies are 
observed for $k>k_d$, in contrast to $E(k)\sim k^{-5}$ for the NSHS simulations we find $E(k)\sim k^{-3}$ for the NSD simulations.

\begin{figure}[!h]
    \includegraphics[width=0.48\textwidth]{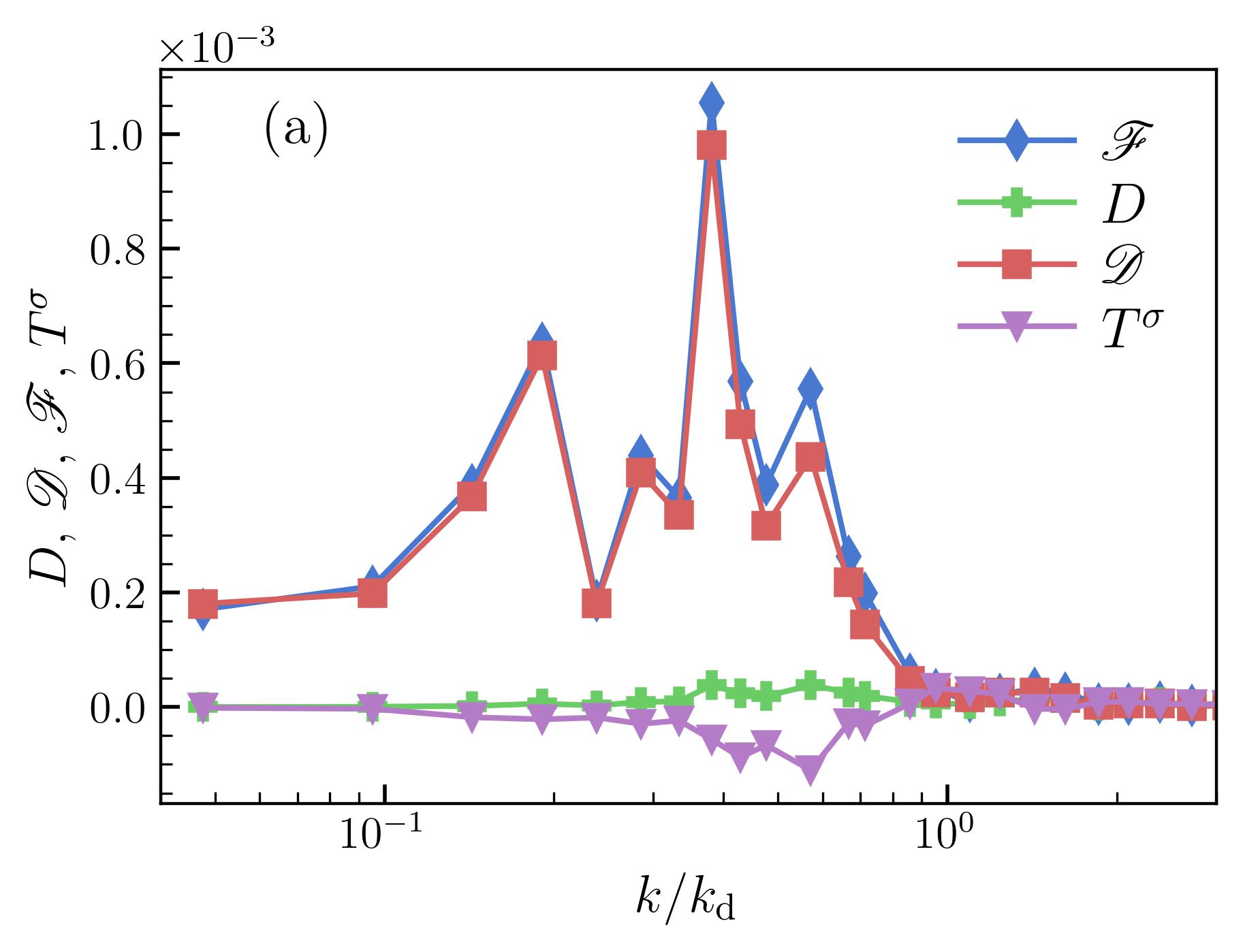}
    \includegraphics[width=0.48\textwidth]{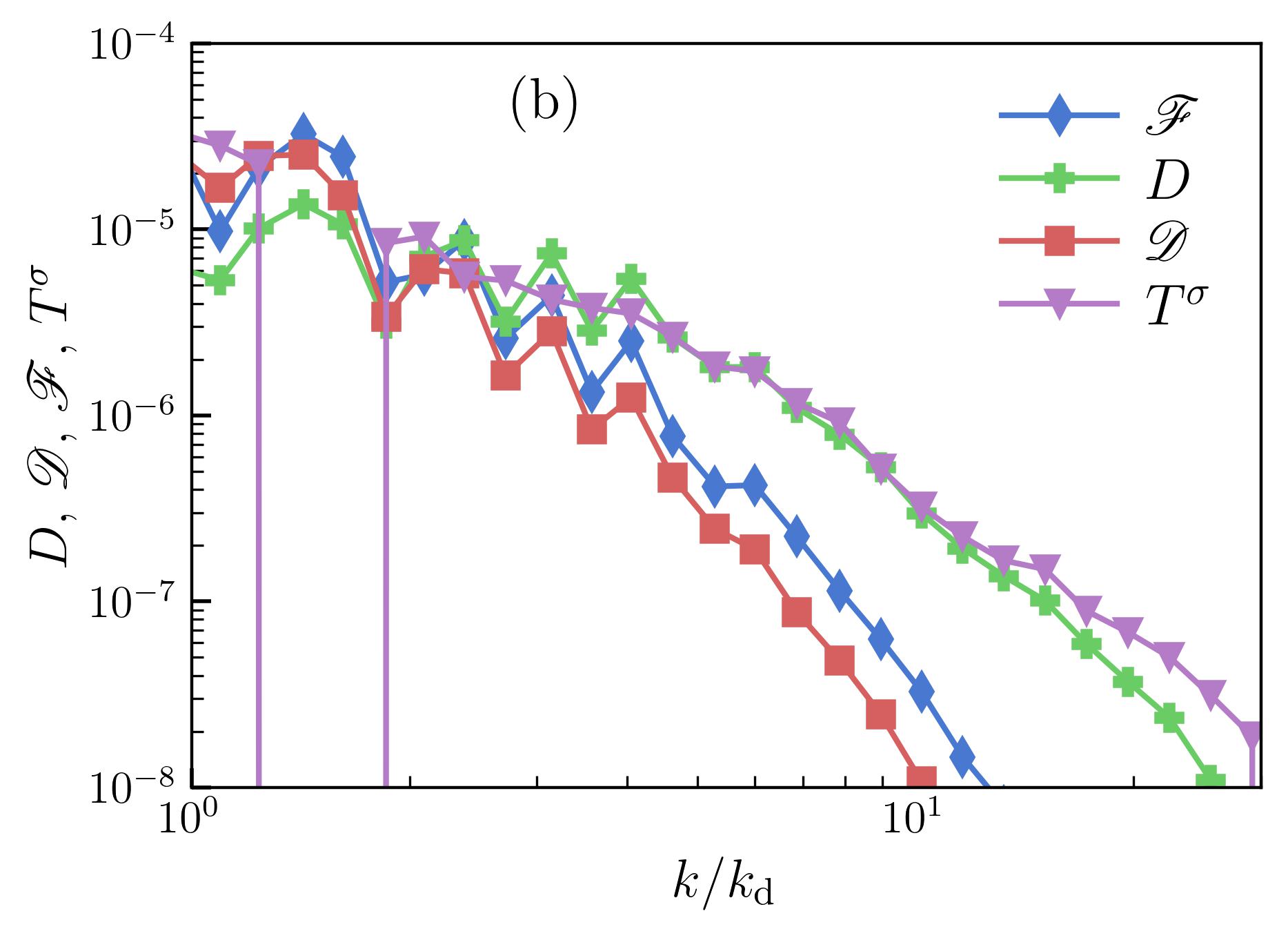}
    \caption{\label{fig:2dbal} Different contributions towards the energy budget for (a) $k<k_d$ and (b) $k>k_d$ obtained from NSD simulation.}
\end{figure}

Using \eqref{eq:drag}, and ignoring the inertial contributions, we obtain the following energy balance 
\begin{align}
{\mathcal F}(k) + T^\sigma(k) = \underbrace{\nu k^2 E(k)}_{\mathcal{D}} + \underbrace{\alpha  E(k)}_{D}.
\label{eq:2dbal}
\end{align}

In Fig.~\ref{fig:2dbal}, we plot the contribution of different terms in  \eqref{eq:2dbal} towards energy balance. 
For $k<k_d$, similar to NSHS, we observe that energy injected by buoyancy is balanced by the linear  drag. However, a different balance appears for $k>k_d$. In contrast to NSHS, a dominant balance is observed 
in the NSD equations. The energy transfer by surface tension to small scales
balances viscous dissipation leading to the observed $E(k)\sim k^{-3}$ scaling in
the energy spectrum. Similar small-scale balance has also been reported in earlier
two-dimensional unbounded bubbly flow simulations \citep{ramadugu2020}.

Therefore, we conclude that although the NSD model captures the large scale dynamics of the Hele-Shaw flow (NSHS), it is unable to correctly capture the small scale physics.

\section{Conclusion}\label{Sec:conclusions}
We have investigated the spectral properties of the two-dimensional bubbly flows
under confinement in a Hele-Shaw setup for experimentally relevant $\Ga$ and $\phi$.
The flow visualization in the steady state is similar to earlier experimental observations \citep{bouche2014}.  The energy spectrum obtained from the gap-averaged velocity field shows  $E(k)\sim k$ for $k<k_d$ and $E(k)\sim k^{-5}$ for $k>k_d$. We also observe an intermediate scaling range with $E(k)\sim k^{-3}$ around $k\sim k_d$. A scale-by-scale energy budget analysis reveals the dominant balances. For $k<k_d$, energy injection balances dissipation due to drag, whereas for $k>k_d$, the net injection balances net dissipation.   Finally, we show that the Navier-Stokes equation with a linear drag can be used to approximate large scale flow properties of bubbly Hele-Shaw flow but it fails to correctly capture energy balance at scales smaller than the bubble diameter.

\section*{Conflict of Interest Statement}
The authors declare that the research was conducted in the absence of any commercial or financial relationships that could be construed as a potential conflict of interest.

\section*{Author Contributions}
P.P contributed to conception and design of the study.  R.R. and V.P. contributed equally. R.R. performed initial NSHS and NSD simulations. V.P. performed NSHS simulations. V.P. and P.P. performed the analysis and wrote the manuscript.  All authors read, and approved the submitted version.

\section*{Funding}
We acknowledge support from  the Department of
Atomic Energy (DAE), India under Project Identification No. RTI 4007, and DST
(India) Project Nos. ECR/2018/001135 and DST/NSM/R\&D\_HPC\_Applications/2021/29.

\REM{
\section*{Acknowledgments}

\section*{Supplemental Data}

\section*{Data Availability Statement}
The datasets [GENERATED/ANALYZED] for this study can be found in the [NAME OF REPOSITORY] [LINK].
}






\end{document}